\documentclass[journal]{IEEEtran}
\usepackage{xcolor,soul,framed} 
\colorlet{shadecolor}{yellow}
\usepackage{graphicx}
\graphicspath{{../pdf/}{../jpeg/}}
\DeclareGraphicsExtensions{.pdf,.jpeg,.png}
\usepackage[cmex10]{amsmath}

\usepackage{url}
\usepackage{bm}
\usepackage{enumerate}
\usepackage{amsfonts}
\usepackage{algorithmic}
\usepackage{algorithm}
\usepackage{nomencl}
\usepackage{etoolbox}
\usepackage{cite}
\usepackage{makecell}
\usepackage{diagbox}
\usepackage{amssymb}

\DeclareMathOperator*{\argmin}{arg\,min}
\newtheorem{remark}{Remark}

\begin{document}

\title{Combining Probabilistic Load Forecasts}

\author{Yi Wang,~\IEEEmembership{Student Member,~IEEE,}
        Ning Zhang,~\IEEEmembership{Senior Member,~IEEE,}
        Yushi Tan,\\
        Tao Hong,~\IEEEmembership{Senior Member,~IEEE,}
        Daniel S. Kirschen,~\IEEEmembership{Fellow,~IEEE,}
        Chongqing Kang,~\IEEEmembership{Fellow,~IEEE}    

\thanks{This work was supported in part by National Key R\&D Program of China (No. 2016YFB0900100) (\textit{Corresponding author: Chongqing Kang})}
\thanks{Y. Wang, N. Zhang, and C. Kang are with the State Key Lab of Power Systems, Dept. of Electrical Engineering, Tsinghua University, Beijing 100084, China. (E-mail: cqkang@tsinghua.edu.cn).}
\thanks{Y. Tan and D. S. Kirschen are with the Department of Electrical Engineering, University of Washington, Seattle, WA 98195-2500, USA.}
\thanks{T. Hong is with  the Energy Production and Infrastructure Center, University of North Carolina at Charlotte, Charlotte, NC 28223, USA.}}
\markboth{Submitted to IEEE Trans. Smart Grid For Peer Review}
{Shell \MakeLowercase{\textit{et al.}}: Bare Demo of IEEEtran.cls for IEEE Journals}
\maketitle

\begin{abstract}
Probabilistic load forecasts provide comprehensive information about future load uncertainties. In recent years, many methodologies and techniques have been proposed for probabilistic load forecasting. Forecast combination, a widely recognized best practice in point forecasting literature, has never been formally adopted to combine probabilistic load forecasts. This paper proposes a constrained quantile regression averaging (CQRA) method to create an improved ensemble from several individual probabilistic forecasts. We formulate the CQRA parameter estimation problem as a linear program with the objective of minimizing the pinball loss, with the constraints that the parameters are nonnegative and summing up to one. We demonstrate the effectiveness of the proposed method using two publicly available datasets, the ISO New England data and Irish smart meter data. Comparing with the best individual probabilistic forecast, the ensemble can reduce the pinball score by 4.39\% on average. The proposed ensemble also demonstrates superior performance over nine other benchmark ensembles. 
\end{abstract}

\begin{IEEEkeywords}
Probabilistic load forecasting, quantile regression, pinball loss function, ensemble method, linear programming, forecasts combination.
\end{IEEEkeywords}
\IEEEpeerreviewmaketitle

\makenomenclature
\renewcommand\nomgroup[1]{%
  \item[\bfseries
  \ifstrequal{#1}{C}{Functions}{%
  \ifstrequal{#1}{A}{Indices}{%
  \ifstrequal{#1}{B}{Sets}{%
  \ifstrequal{#1}{D}{Variables}{}}}}%
]}

\mbox{}

\nomenclature[A]{$n$}{Index of probabilistic forecasting methods}
\nomenclature[A]{$q$}{Index of quantiles}
\nomenclature[A]{$t$}{Index of time periods}

\nomenclature[B]{$Q$}{Set of quantiles}
\nomenclature[B]{$T$}{Set of time periods}
\nomenclature[B]{$N$}{Set of forecasting methods}

\nomenclature[C]{$f_{n,q}(\cdot)$}{The $n$-th model for the $q$-th quantile}
\nomenclature[C]{$f_n(\cdot)$}{The $n$-th point forecast model}
\nomenclature[C]{$f_{e,q}(\cdot)$}{The ensemble model for the $q$-th quantile}
\nomenclature[D]{$\bm{\omega}$}{Vector form of the weights of the base methods}
\nomenclature[D]{$\bm{\omega}_q$}{Vector form of the weights for the $q$-th quantile}
\nomenclature[D]{$\omega_n$}{Weight of the $n$-th method}
\nomenclature[D]{$\omega_{n,q}$}{Weight of the $n$-th method for the $q$-th quantile}
\nomenclature[D]{$\hat{y}_{n,t,q}$}{The forecasted $q$-th quantile of the $n$-th method at time $t$}
\nomenclature[D]{$\hat{y}_{t,q}$}{The forecasted $q$-th quantile of the ensemble method at time $t$}
\nomenclature[D]{$L_{n,t,q}$}{Pinball loss of the $n$-th method at time $t$ for the $q$-th quantile}
\nomenclature[D]{$L_{t,q}$}{Pinball loss of the ensemble method at time $t$ for the $q$-th quantile}
\nomenclature[D]{$L$}{Total pinball loss of the ensemble method}
\nomenclature[D]{$L_{n}$}{Total pinball loss of the $n$-th method}
\nomenclature[D]{$y_{t}$}{Actual load at time $t$}
\nomenclature[D]{$\textbf{S}_{t}$}{Sorted quantiles at time $t$}
\nomenclature[D]{$\bm{\beta}_{Aq}$}{Parameters of (C)QRA\_A model for the $q$-th quantile}
\nomenclature[D]{$\bm{\beta}_{Tq}$}{Parameters of (C)QRA\_T model for the $q$-th quantile}
\nomenclature[D]{$\bm{\beta}_{Eq}$}{Parameters of (C)QRA\_E model for the $q$-th quantile}
\printnomenclature

\section{Introduction}
\IEEEPARstart{L}{oad} forecasting is the basis of power system planning and operation. More accurate forecasts help reduce costs and optimize decisions. A traditional load forecasting process generates only a single value as the estimate of future load for a given timestamp. However, uncertainty on the demand side has been drastically increasing during the recent years. For example, mid- or long-term loads are significantly influenced by economic development and renewable energy deployment, both of which are highly uncertain \cite{ruiz2018voltage}. In a short horizon, the loads are strongly affected by the energy production from renewable energy sources and operations of storage devices. The volatility of the load in a small area or from a single household is much greater than that of the aggregate load at city or state level. Probabilistic forecasts, in the form of intervals, densities, or quantiles, can provide more comprehensive information about uncertainties of the future load than single-valued forecasts \cite{hong2016probabilistic1}.

Producing quantile forecasts was the theme of the Global Energy Forecasting Competition 2014 (GEFCom2014) \cite{hong2016probabilistic}. Gailand \textit{et al}. took the top place of the load forecasting track of GEFCom2014 using a quantile generalized additive model (quantGAM), a hybrid of quantile regression and generalized additive models \cite{gaillard2016additive}. The second top team, Dordonnat \textit{et al}. first developed a point forecasting model based on semi-parametric regression, and then fed the model with different temperature scenarios to generate probabilistic forecasts \cite{dordonnat2016gefcom2014}. The third winning team, Jingrui Xie, took a similar strategy by feeding temperature scenarios to a multiple linear regression based point forecasting model \cite{xie2016gefcom2014}. Xie and Hong further compared three temperature scenario generation methods for probabilistic load forecasting in \cite{xie2016temperature}. Researchers have also developed other means to generate probabilistic forecasts, such as residual simulation~\cite{7163624}, combining point forecasts~\cite{7137662}, and using probabilistic forecasting techniques such as quantile regression~\cite{taieb2016forecasting,wangembedding}. Some works about probabilistic forecasting consider the integration of renewable energy integration~\cite{8066371,ali2018hybrid}. A more comprehensive review of probabilistic load forecasting can be found in \cite{hong2016probabilistic1}.

Probabilistic load forecasting has also been applied to individual loads at household or building level to capture uncertainty. Arora and Taylor captured the uncertainties on individual residential load profiles using both kernel density and conditional kernel density estimation methods~\cite{arora2016forecasting}. They also demonstrated that conditional kernel density can capture the seasonality and describe kernel bandwidth selection and boundary correction methods. Taieb \textit{et al}.~\cite{taieb2016forecasting} combined gradient boosting and additive models in a single quantile regression method. Kou and Gao~\cite{kou2014sparse} proposed a sparse heteroscedastic forecasting method based on a Gaussian process for day-ahead forecasting in energy-intensive enterprises. 

It is widely agreed that no individual forecasting method is the best for all datasets. Combining different forecasts usually reduces the overall risk of making a poor model selection. Forecast combination or ensemble methods can be classified into homogeneous ensemble methods and heterogeneous ensemble methods \cite{mendes2012ensemble}. The former uses the same algorithm with diverse input data, input features, or output targets, while the latter combines several forecasts with the hope that diversity can help improve the results. 

Both methods have been adopted in point load forecasting. Nowotarski \textit{et al}. \cite{nowotarski2016improving} took the homogeneous method, showing that ensembles of 8 sister load forecasts outperformed the best individual point forecasts. Li \textit{et al}.~\cite{li2016novel} developed a homogeneous ensemble method by varying the mother wavelet and the number of decompositions of the wavelet transformation. They then combined the forecasts using partial least squares regression. At NPower Forecasting Challenge 2015, the BigDEAL team, who won a top 3 place, used a heterogeneous ensemble by combining independent forecasts from various techniques, such as multiple linear regression, autoregressive integrated moving average, artificial neural networks, and random forecasts \cite{xie2015combining}. Dudek combined 10 forecasts from different models, including $k$-means, exponential smoothing, and neural networks using simple averaging and weighted averaging, where the weights were determined by the performance of the individual models~\cite{dudek2016heterogeneous}.


Some combination methods can also be applied to different individual probabilistic forecasts to produce another probabilistic forecast. 
Hall and Mitchell~\cite{hall2007combining} chose to minimize the Kullback-Leibler distance between the forecast and true but unknown densities. Clements and Harvey~\cite{clements2011combining} guided the search for optimal weights  with the logarithmic probability score (LPS) and use the maximum likelihood (ML) method to solve the weights. Jore \textit{et al}.~\cite{jore2010combining} used recursive weights to combine the vector autoregressive (VAR)- and autoregressive (AR)-based density forecasts. Combining probabilistic forecasts is still challenging because the problem is usually formulated as a nonlinear and non-convex optimization problem, so that global optimality cannot be guaranteed and the combined results may be worse than individual forecasts.

The literature of combining probabilistic load forecasts is still quite limited. Mangalova and Shesterneva \cite{mangalova2016sequence} used Nadaraya-Watson estimators to combine multiple density forecasts in a sequential fashion, which is similar to the procedure of model generation in the bagging-based ensemble method. Haben and Giasemidis \cite{haben2016hybrid} used different weights for the kernel density estimation method and the quantile regression methods to achieve minimum pinball scores. Nevertheless, none of them addressed how to determine the optimal weights for more than two  probabilistic forecasts. 


Since many probabilistic forecasts characterize future uncertainties in the form of quantiles, combining these quantile forecasts to improve the skill of the ensemble would be quite meaningful in practice. In this paper, \textit{we contribute to the probabilistic load forecasting literature by proposing a constrained quantile regression averaging (CQRA) method for quantile forecast combination}. For parameter estimation, we formulate the proposed CQRA method as a linear programming (LP) problem that minimizes the pinball loss, of which the solution includes the optimal weights for the individual probabilistic forecasts. 
We construct case studies based on two publicly available datasets: the zonal and system level loads from ISO New England (ISO-NE) and the household-level loads from the Commission for Energy Regulation (CER) in Ireland. When using the proposed method to combine forecasts at the target quantile, we can obtain an ensemble that outperforms all individual probabilistic forecasts as well as nine other benchmarks most of the time.

The rest of this paper is organized as follows.
Section~\ref{problem} introduces the probabilistic load forecast combination problem and identifies several key challenges. Section~\ref{methods} introduces the techniques that we are going to use to generate individual probabilistic forecasts. Section~\ref{ensemble} derives an LP formulation of the probabilistic forecast combination problem. Section~\ref{evalutaiton} introduces evaluation criteria to quantify the performance of probabilistic forecasts and lists several benchmarks for comparison. Section~\ref{casestudy} presents the case study settings, forecast results and comparison. Section~\ref{conclusion} conclusions the paper with an outlook of future research.

\section{Problem Statement}
\label{problem}
Forecasting combination refers to the integration of a series of forecasting models to formulate the final forecasting model \cite{mendes2012ensemble}.
The combination process can be divided into two stages: 1) generating a set of forecasting models; 2) integrate these models.
This paper endeavors to combine various probabilistic forecasts to provide a improved quantile forecast. Two issues corresponding to the two processes should be addressed:
\begin{enumerate}
\item Model generating: how to generate different quantile probabilistic forecasts with higher diversity about future uncertainties? 
\item Model ensemble: how to determine the weights of the individual models to achieve the optimal combination?
\end{enumerate}

The first issue will be addressed in Section \ref{methods} by using three typical quantile regression models, and the second issue will be addressed in Section \ref{ensemble} by formulating an optimization problem.

\section{Probabilistic Load Forecasting\\ Model Generation}
\label{methods}
In this section, both homogeneous and heterogeneous methods are used to obtain various basic quantile probabilistic load forecasting models. Density or distribution forecast results can also be converted to the form of quantiles. Since the application of basic forecasting models is not the main consideration of this paper, we only introduce three typical regression models for quantile probabilistic forecasts. We obtain different forecasts by varying the hyperparameters and input data. 

Quantile regression itself can be formulated as an optimization problem to minimize the pinball loss. The pinball loss is a comprehensive index to evaluate the reliability, sharpness, and calibration of the forecasts. 
It is defined for any quantile $q \in (0, 1)$ through a weighted absolute error:
\begin{align}
\label{eq:1}
L_{n,t,q}({{\hat{y}}_{n,t,q}},{{y}_{t}})=\Bigg\{
\begin{aligned}
&(1-q)({{{\hat{y}}}_{n,t,q}}-{{y}_{t}})\text{  }{{{\hat{y}}}_{n,t,q}}\ge {{y}_{t}} \\
&q({{y}_{t}}-{{{\hat{y}}}_{n,t,q}})\text{        }\text{        }\text{        }\text{        }\text{        }\text{        }\text{        }\text{        }{{{\hat{y}}}_{n,t,q}}<{{y}_{t}}.
\end{aligned}
\end{align}

For a certain regression model, its parameters can be optimized by the following optimization problem: 
\begin{equation} \label{eq:2}
\begin{aligned}
\textbf{W}_{n,q}=\arg \underset{\textbf{W}_{n,q}}{\mathop{\min }}\,\sum_{t\in T}\sum_{q\in Q}L_{n,t,q}\big(f_{n,q}(\textbf{X}_{n,t},\textbf{W}_{n,q}),{y}_{t}\big).
\end{aligned}
\end{equation}
\nomenclature[D]{$\textbf{W}_{n,q}$}{Parameters of the $n$-th model for the $q$-th quantile}
\nomenclature[D]{$\textbf{X}_{n,t}$}{Inputs of the $n$-th model at time $t$}
The regression functions $f_{n,q}(\cdot)$ are distinct for different quantile regression models, such as artificial neural network (ANN), gradient boosting regression tree (GBRT), and random forests (RF). The integration of these regression methods and the pinball loss function in load forecasting is briefly introduced in the following paragraphs.

\subsection{Quantile Regression Neural Network (QRNN)}
For linear regression, the parameters $\textbf{W}_{n,q}$ include the slopes $\bm{\beta}_{n,q}$ and intercept $b_{n,q}$:
\begin{equation} \label{eq:3}
\begin{aligned}
\hat{y}_{t}=f_n(\textbf{X}_{n,t},\textbf{W}_{n,q})=\bm{\beta}_{n,q}\textbf{X}_{n,t}+b_{n,q}.
\end{aligned}
\end{equation}

Similarly, for ANN with a hidden layer, the parameters $\textbf{W}_{n,q}$ include the hidden layer weights $\textbf{W}_{n,q}^{(h)}$, the hidden layer basis $\textbf{b}_{n,q}^{(h)}$, the output weights $\textbf{W}_{n,q}^{(o)}$, and the output biases $b_{n,q}^{(o)}$. Thus, the output of the hidden layer $\textbf{y}_{t}^{(h)}$ can be calculated as follows:
\begin{equation} \label{eq:4}
\begin{aligned}
\textbf{y}_{t}^{(h)}=g_n^{(h)}(\textbf{W}_{n,q}^{(h)}\textbf{X}_{n,t}+\textbf{b}_{n,q}^{(h)}),
\end{aligned}
\end{equation}
where $g_n^{(h)}(\cdot)$ denotes the activation function of the hidden layer. Choices of activation function include sigmoid function, tanh function, and rectified linear unit (ReLU). We choose the sigmoid function in this paper.

The final output, i.e., the forecast value, is:
\begin{equation} \label{eq:5}
\begin{aligned}
\hat{y}_{t}=g_n^{(o)}(\textbf{W}_{n,q}^{(o)}\textbf{y}_{t}^{(h)}+b_{n,q}^{(o)}),
\end{aligned}
\end{equation}
where $g_n^{(o)}(\cdot)$ denotes the activation function of the output layer.

To reduce the risk of overfitting the ANN, a weight decay regularization term that penalizes large weights is added to the total pinball loss in Eq. \eqref{eq:2}.
\begin{equation} \label{eq:6}
\begin{aligned}
\textbf{W}_{n,q}=\arg \underset{\textbf{W}_{n,q}}{\mathop{\min }}\,\sum_{t\in T}E\{L_{n,t,q}(\hat{y}_{t},{y}_{t})\}+\lambda \|\textbf{W}_{n,q}^{(h)}\| ,
\end{aligned}
\end{equation}
where $\lambda$ is a positive constant and represents the proportion of the weight decay term.

For traditional ANN with a mean square error (MSE) loss function, which is differentiable everywhere, the back-propagation algorithm can be used to search the weights and basis. For a QRNN with a pinball loss function, which is only partially differentiable, an approximation function that is differentiable everywhere can be used to replace the pinball loss function \cite{chen2007finite}. The number of neurons in the input layer is determined by the dimensions of the input data, and a different number of neurons in the hidden layer results in a different QRNN model. This number is set to 4, 5, 6, 7, and 8 to train five different QRNN models for the ensemble process.

\subsection{Quantile Regression Random Forests (QRRF)}
Random Forests (RF) is a decision tree (classification and regression tree, CART)-based learning method. Each decision tree is trained using a bootstrap sampled subset of the original training dataset. $N_{RF}$ denotes the number of trees to grow and $M_{RF}$ the number of variables randomly sampled as candidates at each split. Thus, the parameters $\textbf{W}_{n,q}$ include random parameter $\textbf{W}_{n,q,n_{RF}}$ for each individual decision tree $T(\textbf{W}_{n,q,n_{RF}})$. The parameters for each tree can be obtained through the $M_{RF}$ training data, where the training method is the same as the CART training method. Each individual decision tree provides a forecast result:
\begin{equation} \label{eq:7}
\begin{aligned}
\hat{y}_{t,n_{RF}}=T(\textbf{W}_{n,q,n_{RF}},\textbf{X}_{n,t}).
\end{aligned}
\end{equation}
Then, the forecast of RF is:
\begin{equation} \label{eq:8}
\begin{aligned}
\hat{y}_{t}=\frac{1}{N_{RF}}\sum_{n_{RF}=1}^{N_{RF}}\hat{y}_{t,n_{RF}}.
\end{aligned}
\end{equation}

A traditional RF attempts to reduce the absolute error by approximating the conditional mean. Because a RF can provide multiple estimates of ${y}_{t}$, a QRRF can be easily implemented by computing the distribution function and quantiles based on these estimates of $\hat{y}_{t,n_{RF}}$. While RF is essentially an ensemble regression method, it is used as a base probabilistic load forecasting model in this paper.

To guarantee that every input row is predicted at least a few times, the number of trees to grow $N_{RF}$ should not be too small. $N_{RF}=500$ is used in this paper. However, the number of variables randomly sampled from dataset $M_{RF}$ is set to $1/2$, $1/3$, $1/4$, and $1/5$ of the size of the whole training dataset to train four different QRRF models for the ensemble process.

\subsection{Quantile Regression Gradient Boosting (QRGB)} 
The gradient boosting decision tree (GBDT) method is also a decision-tree-based ensemble regression method. In contrast to the RF, GBDT adds a new decision tree to fit the residuals from previously generated trees. The final regression model is trained in an iterative manner. We denote $M_{GB}$ as the total number of terminal nodes of the learned tree and $N_{GB}$ as the number of iterations.  Thus, the parameters $\textbf{W}_{n,q}$ include a random parameter $\textbf{W}_{n,q,n_{GB}}$ for each individual decision tree $f_{n,n_{GB}}(\textbf{W}_{n,q,n_{GB}})$. For the $n_{GB}$-th iteration,
\begin{equation} \label{eq:9}
\begin{aligned}
f_{n}^{n_{GB}}(\textbf{X}_t)=f_n^{n_{GB}-1}(\textbf{X}_{n,t})+\\\lambda\sum_{m_{GB}=1}^{M_{GB}}\gamma_{m_{GB},n_{GB}}1(\textbf{X}_{n,t}\in R_{m_{GB},n_{GB}}),
\end{aligned}
\end{equation}
where $\lambda$ denotes the learning rate of the gradient boosting tree; $\{R_{m_{GB},n_{GB}}\}_1^{M_{GB}}$ denotes the $M_{GB}$ disjoint regions at the $n_{GB}$-th iteration; $\gamma_{m_{GB},n_{GB}}$ denotes the optimal terminal node estimates and can be calculated as follows:
\begin{equation} \label{eq:10}
\begin{aligned}
\gamma_{m_{GB},n_{GB}}=\arg \underset{\gamma}{{\min }}\,\sum_{t\in T}\\L_{n,t,q}(f_n^{n_{GB}-1}(\textbf{X}_{n,t},\textbf{W}_{n,q,n_{GB}-1})+\gamma,{y}_{t}).
\end{aligned}
\end{equation}

In this way, each iteration attempts to minimize the previous residuals. More details about the derivation of GBDT can be found in \cite{friedman2002stochastic}. One of the main hyperparameters of the QRGB is the number of iterations, i.e., the total number of trees to fit, $N_{GB}$. We vary $N_{GB}$ from 70 to 100 with an interval of 10 to obtain four QRGB models for the ensemble process.

\subsection{Summary} 
A total of thirteen quantile regression models are generated:
\begin{enumerate}
\item Five QRNN models with 4, 5, 6, 7, or 8 neurons in the hidden layer;
\item Four QRRF models where the number of variables randomly sampled from $M_{RF}$ is set to $1/2$, $1/3$, $1/4$, or $1/5$ of the whole training dataset;
\item Four QRGB models where the number of trees to fit, $N_{GB}$, is set to 70, 80, 90, or 100.
\end{enumerate}

These thirteen quantile regression models, both homogeneous models with different hyperparameters and heterogeneous models, including ANN, RF, and GBDT, constitute a comprehensive model set.

\section{CQRA based Model Combination}
\label{ensemble}
Different indices can be used to quantify the performance of probabilistic load forecasting. For example, reliability and sharpness describe how close and tightly the predicted distribution is to the actual one, respectively; resolution describes how much the predicted interval varies over time \cite{hong2016probabilistic1}. Instead of focusing on a  single aspect, pinball loss  provides a comprehensive score for the performance of probabilistic forecasts and is consistent with the objective function of quantile regression. Thus, the pinball loss is used as a guide for the integration of multiple models.

\subsection{Problem Formulation}
If the weights of each individual model for all the quantiles are identical, the forecast combination model for quantile $q$ can be formulated as follows:

\begin{equation} \label{eq:11}
\begin{aligned}
f_{e,q}(\textbf{X}_{n,t},\bm{\omega})=\sum_{n=1}^N\omega_nf_{n,q}(\textbf{X}_{n,t},\textbf{W}_{n,q}).
\end{aligned}
\end{equation}
The weights $\bm{\omega}$ are estimated by solving the following optimization problem:
\begin{equation} \label{eq:12}
\begin{aligned}
\hat{\bm{\omega}} = &\underset{\bm{\omega}}{\mathop{\argmin }} &&\sum_{q\in Q}\sum_{t\in T}L_{n,t,q}\big(\sum_{n=1}^N\omega_nf_{n,q}(\textbf{X}_{n,t},\textbf{W}_{n,q}),{y}_{t}\big)\\
&s.t.&&\sum_{n=1}^N\omega_n=1\\
&  &&\omega_n \geq 0, ~~\forall n \in \{ 1, \cdots, N \}
\end{aligned}
\end{equation}

By contrast, if the weights of each individual model for the different quantiles are different, the forecast combination method is formulated as:
\begin{equation} \label{eq:13}
\begin{aligned}
f_{e,q}(\textbf{X}_{n,t},\bm{\omega}_q)=\sum_{n=1}^N\omega_{n,q}f_{n,q}(\textbf{X}_{n,t},\textbf{W}_{n,q}).
\end{aligned}
\end{equation}

In the following, we focus on the forecast combination method for each quantile. For each quantile, the determination of weights $\bm{\omega}$ is transformed into solving $Q$ individual optimization problems. The $q$-th problem is: 
\begin{equation} \label{eq:14}
\begin{aligned}
\hat{\bm{\omega}_q}= &\underset{\bm{\omega}_q}{\mathop{\argmin }} &&\sum_{t\in T}L_{n,t,q}(\sum_{n=1}^N\omega_{n,q}f_{n,q}(\textbf{X}_{n,t},\textbf{W}_{n,q}),{y}_{t})\\
&s.t. && \sum_{n=1}^N\omega_{n,q}=1,\\
& &&\omega_{n,q} \geq 0, ~~\forall n \in \{ 1, \cdots, N \}.
\end{aligned}
\end{equation}

\subsection{Data Splitting}
\begin{figure}[h]
\centering
  \includegraphics[width=7cm]{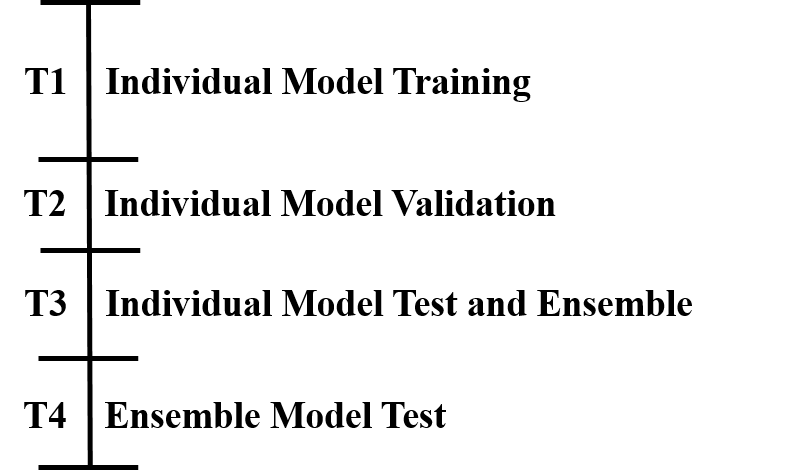}
\caption{Time series splitting for model training, validation, integration, and testing.}
\label{fig:DataSplit}       
\end{figure}

The na\"ive models shown in \eqref{eq:12} and \eqref{eq:14} have a high risk of overfitting. We split a validation dataset from the original training dataset to reduce the overfitting risk. 
 Thus, the whole dataset is divided into four parts, as shown in Fig. \ref{fig:DataSplit}. The first part T1 is used to train each of the models. The second part T2 is used to validate each model for hyperparameter tuning. The third part T3 is used to test each individual model, and the forecast results are used for the model combination in \eqref{eq:12} or \eqref{eq:14} such that the overfitting risk is reduced. 
The last part T4 is used to test the final combined forecasts. Since the solution methods for \eqref{eq:12} and \eqref{eq:14} would be the same, in the following, the algorithms for model integration are described by taking \eqref{eq:12} as an example.  

How to optimally split the time series for model training, validation, integration, and testing is beyond the scope of this paper, and therefore the time series splitting in this paper is roughly conducted based on the number of years. The optimal data split issue has been studied in \cite{larsen1999optimal} and \cite{morrison2011optimal}. In \cite{larsen1999optimal}, the split ratio for model training and validation was investigated. In \cite{morrison2011optimal}, a cross-validation inspired methodology to partition the data into calibration and validation sets was proposed which considers all possible data split ways first and then choose the optimal one. For our problem, the data split is to determine three time breakpoints for T1/T2, T2/T3, and T3/T4. Grid search can be probably applied to to search the optimal data split. We would like to leave it for our future work.

\subsection{Model Combination} 
In this subsection, an algorithm to solve the optimization problem is proposed for model combination.

Since with a limited number of quantiles we cannot obtain a complete distribution, a certain quantile of the weighted sum of several quantile-based distributions cannot be calculated either. In this paper, as proposed in~\cite{granger1989combining}, we simply estimate the $q$-th quantile of the weighted sum of several distributions by the weighted sum of the $q$-th quantiles of all the distributions:
\begin{equation} \label{eq:15}
\begin{aligned}
\hat{y}_{t,q}\approx\sum_{n\in N}\omega_{n,q}\hat{y}_{n,t,q}.
\end{aligned}
\end{equation}

Thus, the loss function in~\eqref{eq:12} can be rewritten as follows:
\begin{equation} \label{eq:16}
\begin{aligned}
\hat{\bm{\omega}}_q&=\arg \underset{\bm{\omega}_q}{\mathop{\min }}\,\sum_{t\in T}L_{t,q}(\hat{y}_{t,q},{y}_{t})\\
&=\arg \underset{\bm{\omega}_q}{\mathop{\min }}\,\sum_{t\in T}\sum_{q\in Q}\max \big\{ q({y}_{t}-\hat{y}_{t,q}),(1-q) (\hat{y}_{t,q}-{y}_{t}) \big\} \\
s.t. \,\; &\hat{y}_{t,q}=\sum_{n\in N}\omega_{n,q}\hat{y}_{n,t,q},~~\sum_{n\in N}\omega_{n,q}=1,~~\omega_n\geq 0 ~~\forall n .
\end{aligned}
\end{equation}

For the $q$-th quantile at time $t$, we introduce auxiliary decision variables $v_{t,q}=\max \big\{ q({y}_{t}-\hat{y}_{t,q}),(1-q) (\hat{y}_{t,q}-{y}_{t}) \big\}$, so the problem in \eqref{eq:16} can be transformed into:
\begin{equation} \label{eq:17}
\begin{aligned}
&\hat{\bm{\omega}_q}=\arg \underset{\bm{\omega}_q}{\mathop{\min }}\,\sum_{t\in T}v_{t,q} \\
s.t. \,\; &\hat{y}_{t,q}=\sum_{n\in N}\omega_{n,q}\hat{y}_{n,t,q},~~\sum_{n\in N}\omega_{n, q}=1,~~\omega_{n,q}\geq 0 ~~\forall n .\\
&v_{t,q}\geq q({y}_{t}-\hat{y}_{t,q}),~~~v_{t,q}\geq (1-q) (\hat{y}_{t,q}-{y}_{t})\\
&\{v_{t,q}-q({y}_{t}-\hat{y}_{t,q})\}\{v_{t,q}-(1-q) (\hat{y}_{t,q}-{y}_{t})\}=0.\\
\end{aligned}
\end{equation}

Without the last constraint in \eqref{eq:17}, the problem is an LP problem. We denote the model without the last constraint in \eqref{eq:17} as the RLP model. Reduction to absurdity could be used to prove that the optimal solution of RLP is also the optimal solution of model~\eqref{eq:17}.

\begin{IEEEproof}
If the optimal solution of RLP $\left[\bm{\omega}, \textbf{v}\right]$ does not satisfy the last constraint, then there exists at least one quantile and time period that satisfies $v_{t,q}> q({y}_{t}-\hat{y}_{t,q})$ and $v_{t,q}> (1-q) (\hat{y}_{t,q}-{y}_{t})$. Then, we can find another value $v'_{t,q}=v_{t,q}-\varepsilon$ that satisfies $v'_{t,q}= q({y}_{t}-\hat{y}_{t,q})>(1-q) (\hat{y}_{t,q}-{y}_{t})$ or $v'_{t,q}=(1-q) (\hat{y}_{t,q}-{y}_{t})> q({y}_{t}-\hat{y}_{t,q})$, where $\varepsilon$ is a positive value. Then, $\left[\bm{\omega}, \textbf{v}'\right]$ instead of $\left[\bm{\omega}, \textbf{v}\right]$ is the optimal solution of RLP.
\end{IEEEproof}

Thus, model \eqref{eq:14} can be transformed into an LP problem with approximation on the combined quantiles with $T$  more variables and $2\times T$ more constraints. 

\begin{remark}
Quantile crossing is a well-known but still not well-addressed issue when quantile regression models fitted separately for different quantiles~\cite{koenker2005quantile}. There are two possible ways to avoid quantile crossing in this context. The first one is to integrate the combination models for individual quantiles into one model. The objective function is to minimize the total pinball losses of all quantiles, with more constraints to avoid quantile crossing. The second one is to conduct naive rearrangement after all quantiles are obtained~\cite{chernozhukov2010quantile, haben2016hybrid}. In this paper, we choose naive rearrangement for three reasons: 1) The individually trained combination models work well in our case studies for both system-level load datasets and residential load datasets with very few time periods existing quantile crossing. The first integration strategy involves large amounts of variables and additional constraints and makes the optimization model much more complex and time-consuming. It might not worth the effort of solving such a complicated model just to hedge against this few quantile crossings; 2) The integration strategy can only guarantee that there is no quantile crossing the ensemble stage. Quantile crossing could still happen in the test and prediction stage. 3) Naive rearrangement has very low  computational burden as well as nice asymptotic properties~\cite{chernozhukov2010quantile}.
\end{remark}

\begin{remark}
Bayesian model averaging (BMA) is another effective way to combine different models to provide probabilistic forecast~\cite{brown2002bayes}. It uses posterior probability as the weight to conduct weighted average to individual models. 
It considers the model uncertainty and combines subjective information, model and data information by setting different prior distributions. The main challenges of the implementation of BMA include model selection, integral computation, individual model posterior probability calculation \cite{hoeting1999bayesian}. Our proposed combining method differs from BMA in different ways: 1) the proposed method determines the weights of individual methods under the guidance of pinball loss instead of the posterior probabilities; 
2) some of the models that we are trying to combine do not have explicit probability structures for calculating likelihood functions. Therefore, the proposed method should be more universal;
3) the proposed method can be formulated as a linear programing problem and is of  computational efficiency and accuracy; whereas posteriors sometimes present multi-modality which may be undesirable in decision making~\cite{faria2008geometric};  4) the proposed formulation can be viewed as a special case of lasso regression~\cite{conflitti2015optimal} and therefore select predictors by enforcing sparsity. However, the candidate models for BMA need to be either carefully pre-screened or adaptively dropped in a heuristic way. 
\end{remark}

\section{Evaluation Indexes and Competing Methods}
In this section, the evaluation index pinball loss is used to quantify the performance of probabilistic load forecasts. In addition, nine different weight determination methods for model ensemble are introduced for comparison in the case studies.
\label{evalutaiton}
\subsection{Evaluation Index}
As stated before, the probabilistic forecasts can be evaluated from different aspects including reliability, sharpness, and resolution \cite{hong2016probabilistic1}. The continuous-ranked probability score (CRPS) is a comprehensive index for the three aspects which is mainly used for density forecasts instead of quantiles. The discrete form of CRPS was proposed in \cite{taieb2016forecasting}. Another index, pinball loss, is another comprehensive index that has been used in many previous studies \cite{hong2016probabilistic1} and global energy forecasting competitions. Thus, we use pinball to evaluate the final forecasts. It is calculated using \eqref{eq:1}. The average of all the quantiles' pinball loss gives the overall performance:
\begin{equation} \label{eq:18}
\begin{aligned}
L=\frac{1}{|T_4|\times Q}\sum_{t\in T_4}\sum_{q\in Q}L_{t,q}(\hat{y}_{t,q},{y}_{t}).
\end{aligned}
\end{equation}
where $|T_4|$ denotes the length of the time period $T_4$.

\subsection{Competing Methods}
The nine competing methods consist of naive sorting method, median value based method, simple averaging, weighted averaging method, three quantile regression averaging (QRA) methods, and two CQRA methods.

\subsubsection{Na\"ive Sorting (NS)}
With each forecasting model producing $Q$ quantiles, a total of $N\times Q$ quantiles can be observed (in some sense) by $N$ forecasting models. By sorting these observations by descending order, a new sequence $\mathbf{S}_t = \{ S_{t,j}, j = [1, Q \times N] \}$ can be obtained. And therefore the $q$-th quantile is estimated as follows:
\begin{equation} \label{eq:20}
\begin{aligned}
\hat{y}_{t,q} = S_{t,1 + (q - 1) N}.
\end{aligned}
\end{equation}

\subsubsection{Median Value (MED)}
The median value of the $N$ $q$-th quantiles is selected as the final quantile:
\begin{equation} \label{eq:20}
\begin{aligned}
\hat{y}_{t,q}=S_{t,1 + (q-1) N+[N/2]}.
\end{aligned}
\end{equation}

\subsubsection{Simple Averaging (SA)}
The simple averaging strategy applies equal weights to different methods:
\begin{equation} \label{eq:20}
\begin{aligned}
w_{n,q}=1/N.
\end{aligned}
\end{equation}

Then, the final combined forecasts are calculated according to Eq. \eqref{eq:15}.

\subsubsection{Weighted Averaging (WA)}
The basic idea of the weighted averaging method is that methods with higher accuracy should be given higher weights:
\begin{equation} \label{eq:21}
\begin{aligned}
w_{n,q}=\frac{\frac{1}{L_{n,q}}}{\sum_{n\in N}\frac{1}{L_{n,q}}}.
\end{aligned}
\end{equation}

Similarly, the final combined forecasts are also calculated according to Eq. \eqref{eq:15}.

\subsubsection{QRA-E}
As stated above, $N$ forecasting models produce $N\times Q$ quantiles. These quantiles can also be viewed as $N\times Q$ point forecasts which are denoted as $\textbf{S}_{At}|_{1\times (Q\times N)}$. Then, we can calculate the average quantiles $\textbf{S}_{Et}|_{1\times N}$, where the element of the average quantiles $\textbf{S}_{Et,n}$ is calculated as follows:
\begin{equation} \label{eq:21}
\begin{aligned}
{S}_{Et,n}=\frac{1}{Q}}{\sum_{i=Q\times (n-1)+1}^{Q\times n}{S}_{At,i}.
\end{aligned}
\end{equation}

QRA-E yields new quantiles by applying linear quantile regression to the average of the $Q$ quantiles $\textbf{S}_{Et}$:
\begin{equation} \label{eq:21}
\begin{aligned}
\hat{y}_{t,q}=\textbf{S}_{Et}\bm{\beta}_{Eq}.
\end{aligned}
\end{equation}
\nomenclature[D]{$\textbf{S}_{At}$}{All quantiles of the $N$ models}
\nomenclature[D]{$\textbf{S}_{Et}$}{Average quantiles of the $N$ models}
\nomenclature[D]{$\textbf{S}_{Tt,q}$}{Targeted $q$-th quantiles of the $N$ models}
The parameters are estimated by minimizing the Pinball loss function:
\begin{equation} \label{eq:21}
\begin{aligned}
\bm{\beta}_{Eq}=\arg \underset{\bm{\beta}_{Eq}}{\mathop{\min }}\,\sum_{t\in T}\sum_{q\in Q}L_{n,t}\big(\textbf{S}_{At}\bm{\beta}_{Eq},{y}_{t}\big).
\end{aligned}
\end{equation}

\subsubsection{QRA-A}
Compared with QRA-E, instead of conducting QRA on the averaged quantiles, QRA-A yields new quantiles by applying linear quantile regression to all quantiles $\textbf{S}_{At}$: 
\begin{equation} \label{eq:21}
\begin{aligned}
\hat{y}_{t,q}=\textbf{S}_{At}\bm{\beta}_{Aq}.
\end{aligned}
\end{equation}

\subsubsection{QRA-T}
We can also select the targeted $q$-quantiles from $\textbf{S}_{At,q}$ for QRA. The targeted quantiles $\textbf{S}_{Tt,q}|_{1\times N}$ are selected as follows:
\begin{equation} \label{eq:21}
\begin{aligned}
{S}_{Tt,q,n}={S}_{At,Q\times (n-1)+q}.
\end{aligned}
\end{equation}

Compared with QRA-A, a slight change of QRA-E is the regressors are the targeted $q$-th quantiles $\textbf{S}_{Tt,q}$:
\begin{equation} \label{eq:21}
\begin{aligned}
\hat{y}_{t,q}=\textbf{S}_{Tt,q}\bm{\beta}_{Tq}.
\end{aligned}
\end{equation}

It should be noted that the quantiles used for QRA are different and of different dimensions. Thus, the dimensions of the parameters (weights) of the regression model $\bm{\beta}_{Eq}$, $\bm{\beta}_{Aq}$, $\bm{\beta}_{Tq}$ are also different. 

\subsubsection{CQRA-E}
Compared with QRA-E, the slight changes of CQRA-E are the added constraints:
\begin{equation} \label{eq:21}
\begin{aligned}
\bm{\beta}_{Eq}=\arg \underset{\bm{\beta}_{Eq}}{\mathop{\min }}\,\sum_{t\in T}\sum_{q\in Q}L_{t,q}\big(\textbf{S}_{Et}\bm{\beta}_{Eq},{y}_{t}\big).\\
s.t.~~~~\sum_{n\in N}\beta_{Eq,n}=1,~~~\beta_{Eq,n}\geq 0 ~~\forall n .
\end{aligned}
\end{equation}

\subsubsection{CQRA-A}
Compared with CQRA-E, CQRA-A conducts constrained regression on all the quantiles $\textbf{S}_{At}$:
\begin{equation} \label{eq:21}
\begin{aligned}
\bm{\beta}_{Aq}=\arg \underset{\bm{\beta}_{Aq}}{\mathop{\min }}\,\sum_{t\in T}\sum_{q\in Q}L_{t,q}\big(\textbf{S}_t\bm{\beta}_{Aq},{y}_{t}\big).\\
s.t.~~~~\sum_{n\in N}\beta_{Aq,n}=1,~~~\beta_{Aq,n}\geq 0 ~~\forall n .
\end{aligned}
\end{equation}

Note that the proposed method in this paper can be called CQRA-T, because the targeted quantiles are used for regressors. The estimated weights for each quantile $\bm{\omega}_q$ is equal to the parameters $\bm{\beta}_{Tq}$ in CQRA-T model.

Table \ref{competing} summarized the competing QRA methods from two aspects: 1) whether there are constraints for the weights of the models; 2) which quantiles have been considered in the QRA method. 
\begin{table}[]
\centering
\caption{Competing QRA methods}
\label{competing}
\begin{tabular}{|c|c|c|}
\hline
\diagbox[width = 8em]{Quantiles}{Constraints} & With Constraints & Without Constraints \\ \hline
Averaged Quantiles & 5) QRA-E         & 8) CQRA-E           \\ \hline
All Quantiles      & 6) QRA-A         & 9) CQRA-A           \\ \hline
Targeted Quantiles & 7) QRA-T         & CQRA-T (Proposed)   \\ \hline
\end{tabular}
\end{table}

\begin{figure*}[h]
\centering
  \includegraphics[width=11cm]{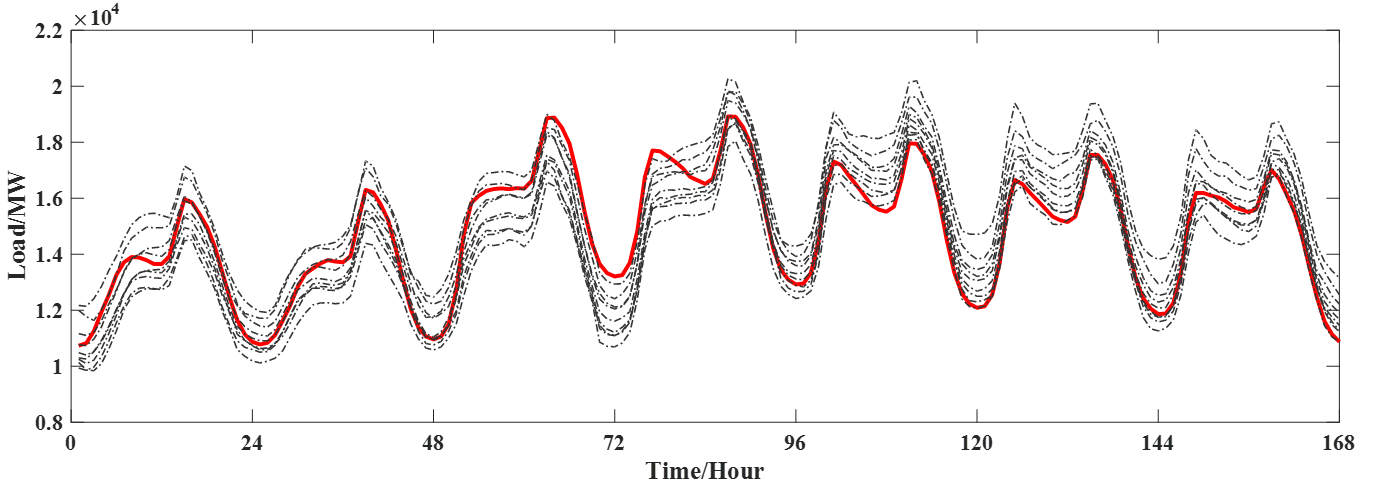}
\caption{Probabilistic load forecasts of the total load (SYS) of one week from Jan 2, 2016 to Jan 8, 2016, where the red line is the real values and the dotted line are forecasted quantiles.}
\label{fig:Quantiles}       
\end{figure*}

\begin{figure*}[h]
\centering
  \includegraphics[width=14cm]{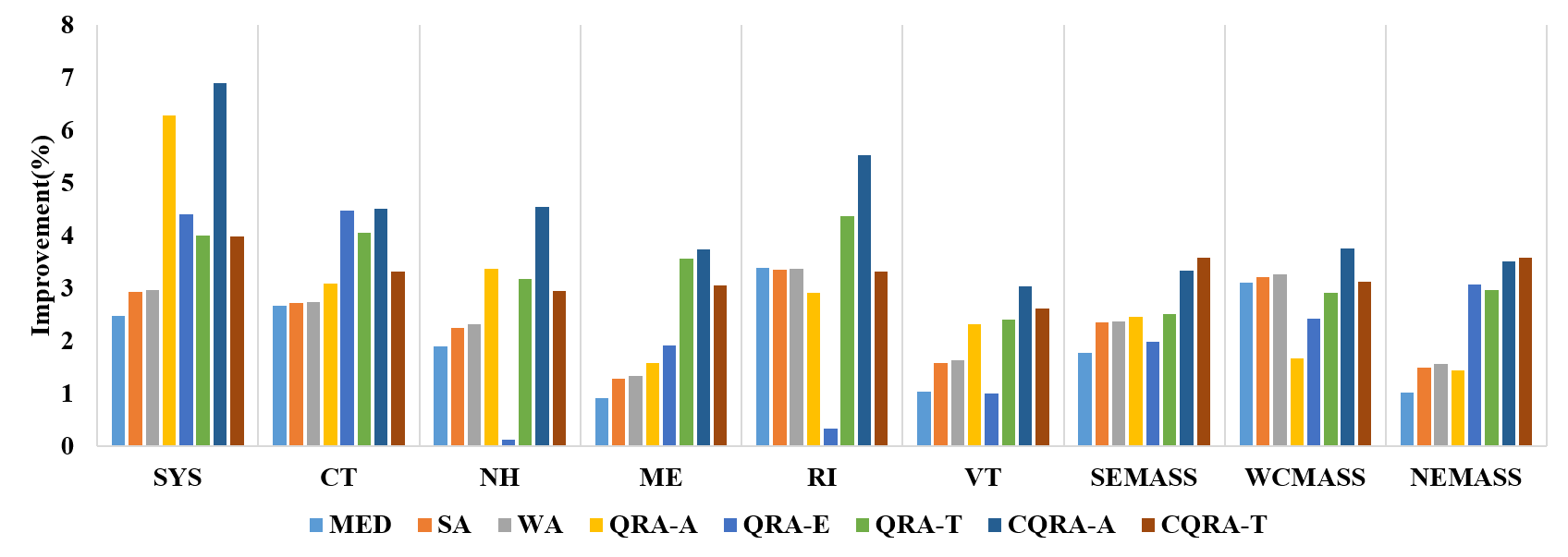}
\caption{Relative improvements of the different combination methods in different zones compared with the best individual (BI) model.}
\label{fig:Improvement}       
\end{figure*}
\section{Case Studies}
\label{casestudy}
We conduct the case studies on both a zonal/system-level load dataset and an individual consumer load dataset on a standard PC with an Intel® Core™ i7-4770MQ CPU running at 2.40 GHz and with 8.0 GB of RAM.
The thirteen quantile regression models are implemented using the qrnn package \cite{cannon2011quantile}, quantregforest package \cite{meinshausen2007quantregforest}, and gbm package \cite{ridgeway2007generalized} in R version 3.3.2. The optimization model for the ensemble is implemented using YALMIP \cite{lofberg2004yalmip} in Matlab R2016a.

\subsection{Day-ahead Zonal/System-Level Load Forecasting}
\subsubsection{Data Description}
The case studies on the zonal/system-level load forecasting are conducted on an ISO-NE load dataset. ISO-NE, an independent regional transmission organization (RTO), serves six New England states: Connecticut (CT), Maine (ME), Massachusetts, New Hampshire (NH), Rhode Island (RI) and Vermont (VT), where Massachusetts consists of three zones: NEMASS, SEMASS, and WCMASS \cite{ISONewEngland}. In this way, ISO-NE is divided into 8 zones. We validate the proposed method on these 8 zones and their total load, which is denoted as SYS. The hourly load data from 2013-1-1 to 2016-12-31, a total of four year data, are used. Specifically, the first two year of the data (T1 \& T2) is used to train and validate the individual models; the third and fourth year of the data are used for the individual model combination (T3), and ensemble model test (T4), respectively. 



\subsubsection{Results}
Fig. \ref{fig:Quantiles} provides the day-ahead probabilistic total (NC) load forecasting results of one week. The red line denotes the actual loads; the gray dotted lines denote the quantiles from 10\% to 90\%  with intervals of 10\%. 

Table \ref{PinballLoss} provides the pinball losses of the different methods for the eight zones and the total load, where the best individual (BI) is the individual model with the best performance. Correspondingly, Fig. \ref{fig:Improvement} shows the relative improvements compared with the BI model. 

It can be seen that most of the combining models (simple averaging, weighted averaging, QRA, and the proposed optimal averaging) outperform the BI model.
There is a special case where the two QRA models for VT has very slightly weak performance compared with BI model. It indicates that the combination method does not always provides better result than the individual methods but would provide improvements with higher credit. In addition, naive sorting has the worst performance because the there might be bias when determining each quantiles according to the sorted sequence. The proposed combining method has the lowest pinball loss for all nine load profiles. The average improvement compared with the best individual is 4.39\% which obviously verify the superiority of the proposed combining forecasts.

We also notice that the addition of the constraints to the method QRA-E (i.e. CQRA-E) so strongly worsen the performance of the combined forecasts. This is because the constraints $\sum_{n\in N}\beta_{Eq,n}=1,~~~\beta_{Eq,n}\geq 0 ~~\forall n$ limit the combination into the a very small interval, the lower and upper bounds are the minimum and maximum excepted quantiles of the $N$ models, respectively. QRA-E is quite similar to traditional QRA which applies quantile regression on the point forecasts. This is also the reason why traditional QRA has not constraints on the weights.

\begin{table*}[]
\centering
\caption{Pinball Losses of the Individual and Combination Methods for Different Zones}
\label{PinballLoss}
\begin{tabular}{|c|c|c|c|c|c|c|c|c|c|}
\hline
\diagbox[width = 8em]{Methods}{Zones}      & SYS               & CT              & NH              & ME              & RI              & VT              & SEMASS          & WCMASS          & NEMASS          \\ \hline
BI    & 288.563          & 81.478          & 27.216          & 18.146          & 21.756          & 12.426          & 42.307          & 41.939          & 63.685          \\ \hline
NS      & 327.569          & 95.058          & 31.586          & 19.003          & 25.738          & 13.247          & 48.817          & 47.041          & 71.873          \\ \hline
MED             & 281.607          & 79.359          & 26.713          & 17.981          & 21.044          & 12.300          & 41.570          & 40.676          & 63.048          \\ \hline
SA   & 280.375          & 79.322          & 26.618          & 17.916          & 21.053          & 12.233          & 41.336          & 40.638          & 62.752          \\ \hline
WA & 280.266          & 79.306          & 26.600          & 17.908          & 21.049          & 12.227          & 41.329          & 40.616          & 62.706          \\ \hline
QRA-E             & 276.417          & 77.995          & 27.184          & 17.806          & 21.683          & 12.303          & 41.484          & 40.949          & 61.793          \\ \hline
QRA-A             & 271.519          & 79.037          & 26.330          & 17.864          & 21.140          & 12.145          & 41.295          & 41.252          & 62.783          \\ \hline
QRA-T            & 277.487          & 78.313          & 26.380          & 17.523          & 20.847          & 12.135          & 41.271          & 40.752          & 61.849          \\ \hline
CQRA-E            & 356.527          & 100.925         & 33.829          & 22.767          & 26.540          & 15.616          & 51.765          & 51.544          & 79.131          \\ \hline
CQRA-A            & 277.510          & 78.870          & 26.437          & 17.610          & 21.059          & 12.109          & \textbf{40.847} & 40.672          & \textbf{61.491} \\ \hline
CQRA-T            & \textbf{269.953} & \textbf{77.961} & \textbf{26.034} & \textbf{17.492} & \textbf{20.619} & \textbf{12.061} & 40.941          & \textbf{40.422} & 61.524          \\ \hline
\end{tabular}
\end{table*}

\begin{figure}[h]
\centering
  \includegraphics[width=9cm]{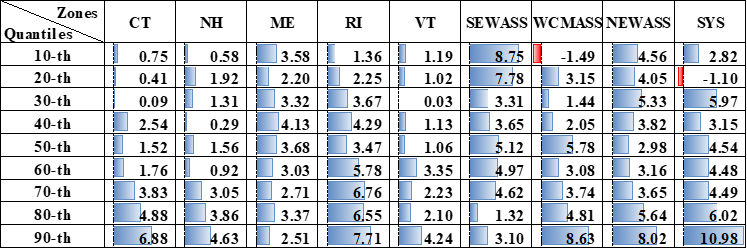}
\caption{Relative improvements (\%) of the different combination methods for different quantiles in different zones}
\label{fig:QuantileImprovement}       
\end{figure}

Fig. \ref{fig:QuantileImprovement} depicts the relative improvements in our proposed method in different zones for different quantiles. The two points marked by the red circle show that not all the quantiles improve. There is no clear relationship between the improvement and the quantile.

\begin{figure}[h]
\centering
  \includegraphics[width=9cm]{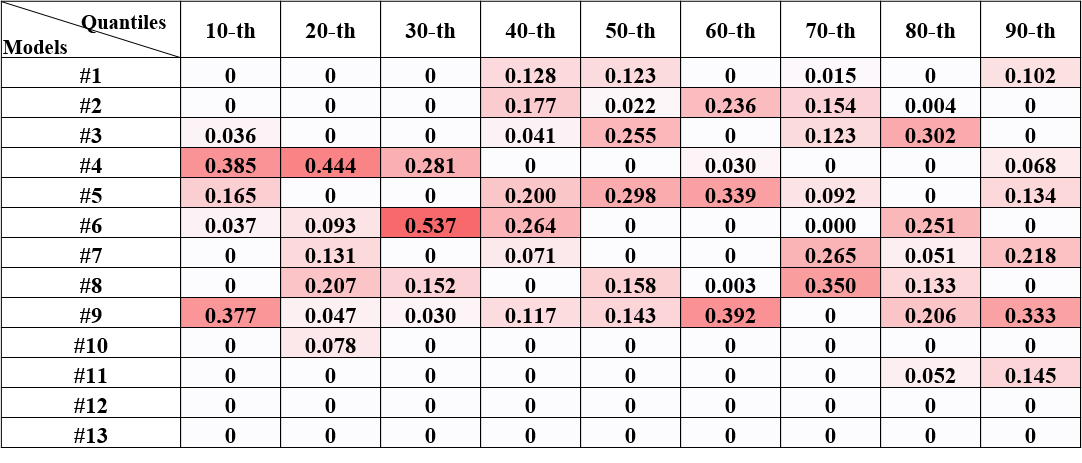}
\caption{Models that are selected for different quantiles for total load (SYS).}
\label{fig:ModelSelection}       
\end{figure}

\begin{figure}[h]
\centering
  \includegraphics[width=9cm]{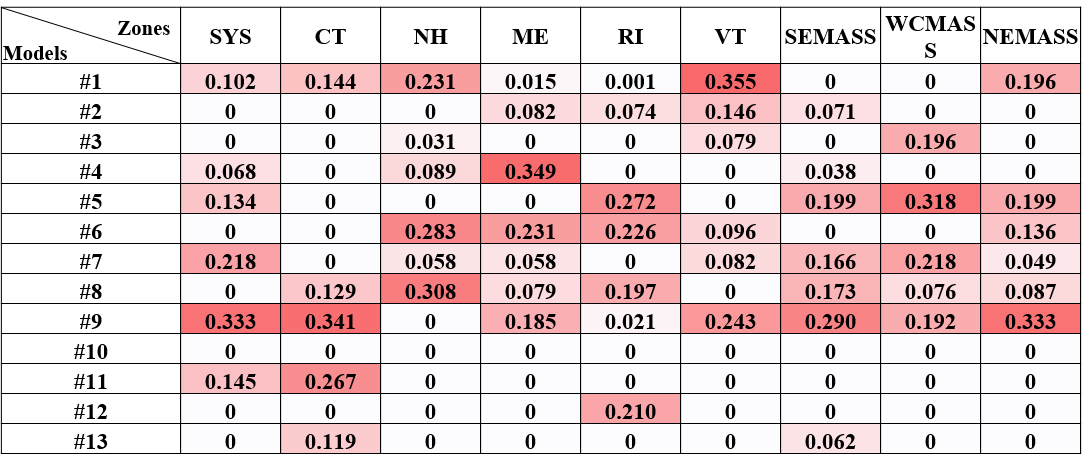}
\caption{Models that are selected for the 90-th quantile for different zones.}
\label{fig:ModelSelection2}       
\end{figure}

In the process of model integration, some models are not shown be used to from the final forecasts. Fig. \ref{fig:ModelSelection} provides  the weights for the different models for the total load (SYS), respectively. Method \#12 and \#13 have been pruned for all quantiles owing to its poor performance. The number of models retained for different quantile ranges from 6 to 9. It should be noted that the pruned model does not necessarily have the worst performance. In addition, the weights for different models does not change smoothly with the change of quantiles. There may exist two reasons: 1) the proposed combining method has constraints on the weights and results in sparsity of the weights; 2) the quantile regression models are trained individually for different individual quantiles and thus the performances of adjunct quantiles do not change smoothly. Fig. \ref{fig:ModelSelection2} gives the weights for the 90-th quantile for different zones in ISO-NE. The results shows that different models have been pruned for the 90-th quantile forecasting for different zones.


\begin{figure*}[h]
\centering
  \includegraphics[width=11cm]{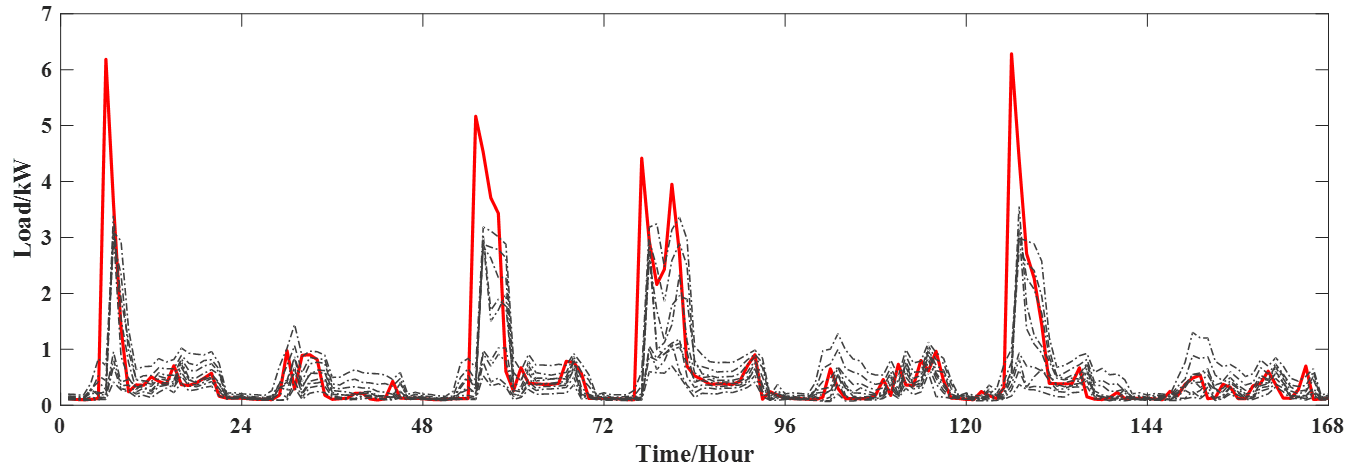}
\caption{Probabilistic load forecasts of consumer \#1002 of one week from July 20, 2010 to July 26, 2010, where the red line is the real values and the dotted line are forecasted quantiles.}
\label{fig:Quantiles2}
\end{figure*}

\subsection{Hour-ahead Residential Load Forecasting} 
\subsubsection{Data Description}
The case study on the individual consumer load forecasting is conducted on a CER Irish dataset, where ten residents are selected. The 30-min data are aggregated to one-hour load profiles from 2009-7-15 to 2010-12-31. The dataset division strategy for the training and testing of individual methods, and the ensemble is the same as that for the ISO-NE dataset.

\begin{table*}[]
\centering
\caption{Pinball Losses of the Individual and Combination Methods for Individual Consumers}
\label{PinballLoss2}
\begin{tabular}{|c|c|c|c|c|c|c|c|c|c|c|}
\hline
\diagbox[width = 8em]{Methods}{Meter ID}   & 1002          & 1003          & 1004          & 1005          & 1008          & 1013          & 1015          & 1016          & 1017          & 1018          \\ \hline
BI    & 0.0548          & 0.1726          & 0.2858          & 0.2215          & 0.2489          & 0.0772          & 0.1071          & \textbf{0.1459} & 0.1214          & 0.1585          \\ \hline
NS      & 0.059           & 0.1949          & 0.3375          & 0.269           & 0.2873          & 0.0871          & 0.1238          & 0.1834          & 0.1351          & 0.1828          \\ \hline
MED             & 0.0542          & 0.1716          & 0.2845          & 0.2236          & 0.2504          & 0.0777          & 0.107           & 0.1539          & 0.1219          & 0.1577          \\ \hline
SA   & 0.054           & 0.1707          & 0.2836          & 0.2211          & 0.2491          & 0.0775          & 0.1067          & 0.1518          & 0.1215          & 0.1573          \\ \hline
WA & 0.054           & 0.1707          & 0.2836          & 0.2211          & 0.249           & 0.0775          & 0.1067          & 0.1518          & 0.1214          & 0.1573          \\ \hline
QRA-E             & 0.0567          & 0.1737          & 0.3031          & 0.234           & 0.259           & 0.0782          & 0.1104          & 0.1562          & 0.1273          & 0.1645          \\ \hline
QRA-A             & 0.0578          & 0.186           & 0.3103          & 0.2612          & 0.2664          & 0.0826          & 0.1138          & 0.1729          & 0.1285          & 0.1686          \\ \hline
QRA-T             & 0.0548          & 0.1709          & 0.288           & 0.2231          & 0.2526          & 0.077           & 0.1081          & 0.1493          & 0.1219          & 0.1576          \\ \hline
CQRA-E            & 0.0696          & 0.2157          & 0.3668          & 0.2787          & 0.317           & 0.0995          & 0.134           & 0.1824          & 0.1586          & 0.1996          \\ \hline
CQRA-A           & 0.0535          & 0.1705          & 0.2876          & 0.2257          & 0.252           & \textbf{0.0766} & 0.1075          & 0.1504          & \textbf{0.1202} & \textbf{0.1565} \\ \hline
CQRA-T            & \textbf{0.0532} & \textbf{0.1689} & \textbf{0.2811} & \textbf{0.2185} & \textbf{0.2458} & 0.0768          & \textbf{0.1064} & 0.1525          & 0.1208          & 0.1574          \\ \hline
\end{tabular}
\end{table*}

\subsubsection{Results}
Fig. \ref{fig:Quantiles2} shows the hour-ahead probabilistic load forecasting results of consumer \#1002. In contrast to the system-level load profiles, the load profile shows great volatility; thus the 10-th and 90-th quantiles do not effectively cover the spikes.

Table~\ref{PinballLoss2} illustrates the pinball losses of the different methods for the 10 selected individual consumers. It can be seen that the proposed combining method has the best performance for all consumers except consumer \#1016. For consumer \#1016, the sample averaging, weighted averaging, QRA, and the proposed averaging do not outperform the best individual method. Again, the results show that the ensemble method does not always have better performance than individual methods. Another interesting observation is that two QRA models have worse performance compared with the best individual for all selected individual consumers. It means that QRA model may not be able to handle high uncertainty load profile such as the individual residential load. However, the proposed combining method is robust for this situation. The essential difference between our proposed combining method and QRA is that the proposed resemble method determines the optimal weights of the most relevant $N$ quantiles while QRA determines the optimal weights of all the $N\times Q$ quantiles. Large number of regressors (quantiles to be regressed) in QRA contain many less relevant regressors and may lead to overfitting. 

\section{Conclusions and Future Work}
\label{conclusion}
This paper proposes a constrained  quantile regression averaging method for quantile forecasting to take full advantage of the increasing number of forecasting models that have been described in the literature. The ensemble problem is formulated as a series of LP problems, each corresponding to a quantile. The results show that the proposed ensemble method effectively improves the forecasting performance in terms of pinball loss compared with individual models. 

Future works will consider two aspects. The first is the investigation of pruning methods such as lasso-based method that can be integrated into the proposed combination method refine the individual probabilistic forecasting model and further improve the performance of the combined model. The second is the extension of our method to probabilistic renewable energy forecasting. Note that our proposed method is performed on quantile results and is not limited to only load forecasting. It is interesting to investigate the improvements of the proposed method on different load datasets and renewable energy forecasting.

\bibliographystyle{IEEEtran}
\bibliography{IEEEabrv,Reference}

\begin{IEEEbiography}[{\includegraphics[width=1in,height=1.25in,clip,keepaspectratio]{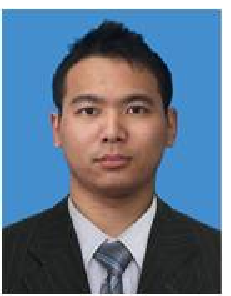}}]{Yi Wang}
(S'14) received the B.S. degree from the Department of Electrical Engineering in Huazhong University of Science and Technology (HUST), Wuhan, China, in 2014.\\
He is currently pursuing Ph.D. degree in Tsinghua University. He is also a visiting student researcher at the University of Washington, Seattle, WA, USA. His research interests include data analytics in smart grid and multiple energy systems.
\end{IEEEbiography}

\begin{IEEEbiography}[{\includegraphics[width=1in,height=1.25in,clip,keepaspectratio]{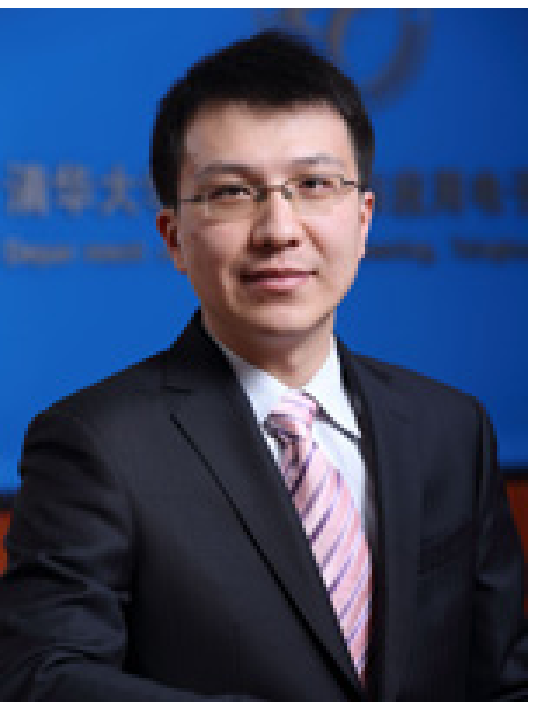}}]{Ning Zhang}
(M'12-SM'18) received both a B.S. and Ph.D. from the Electrical Engineering Department of Tsinghua University in China in 2007 and 2012, respectively.\\
He is currently an Associate Professor at the same university. His research interests include multiple energy systems integration, renewable energy, and power system planning and operation.
\end{IEEEbiography}

\begin{IEEEbiography}[{\includegraphics[width=1in,height=1.25in,clip,keepaspectratio]{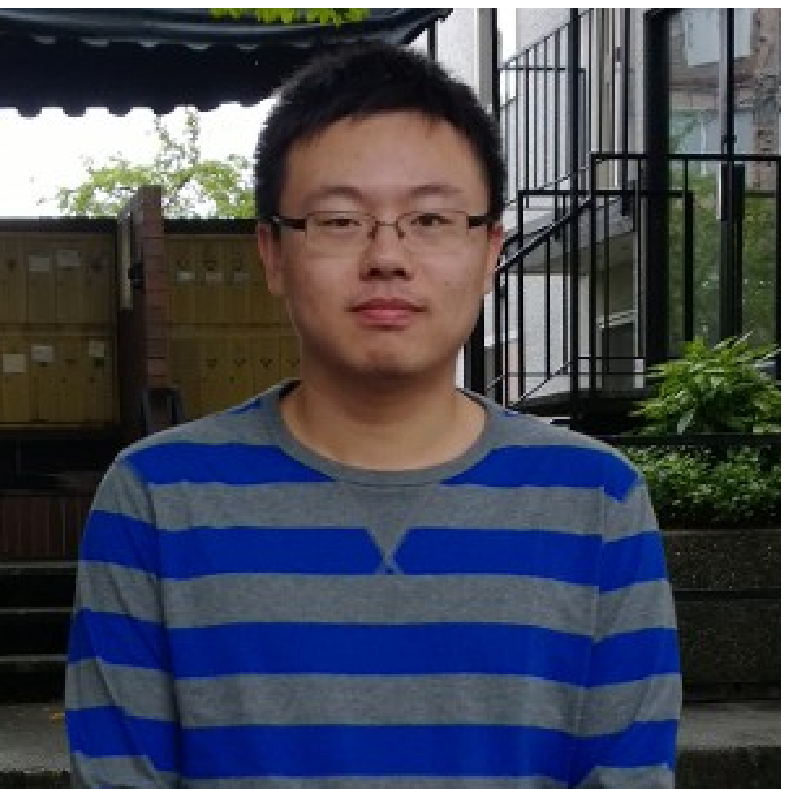}}]{Yushi Tan}
received the B.S. degree from the Electrical Engineering Department of Tsinghua University in China, in 2013.\\
He is currently pursuing Ph.D. degree in the University of Washington, Seattle, WA, USA. His research interests include load forecasting and power system resilience.
\end{IEEEbiography}

\begin{IEEEbiography}[{\includegraphics[width=1in,height=1.25in,clip,keepaspectratio]{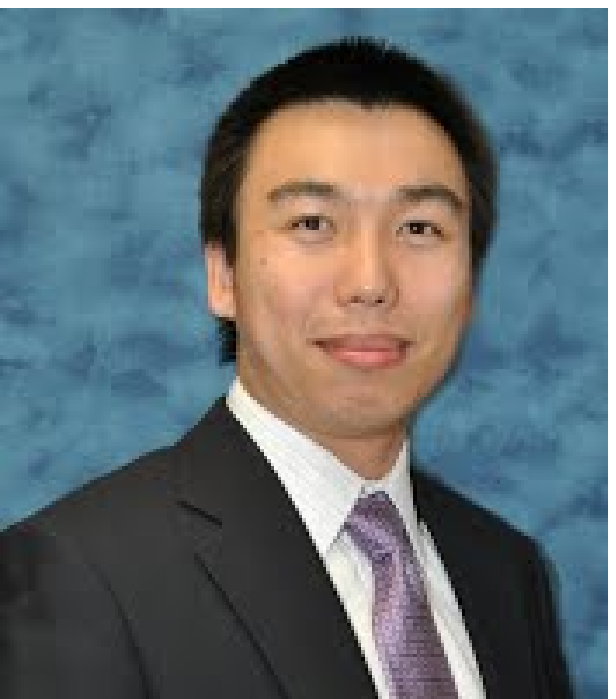}}]{Tao Hong}
received the B.Eng. degree in automation from Tsinghua University, Beijing, China, in 2005, and the Ph.D. degree in operation research and electrical engineering from North Carolina State University, Raleigh, NC, USA, in 2010. \\
He is the Director of the Big Data Energy Analytics Laboratory, the Graduate Program Director, an Assistant Professor of System Engineering and Engineering Management, and an Associate of the Energy Production and Infrastructure Center with the University of North Carolina at Charlotte, Charlotte, NC. He has authored the book Electric Load Forecasting: Fundamentals and Best Practices (OTexts, 2015), and a blog Energy Forecasting since 2013. Dr. Hong is the Founding Chair of the IEEE Working Group on Energy Forecasting and the General Chair of the Global Energy Forecasting Competition.
\end{IEEEbiography}

\begin{IEEEbiography}[{\includegraphics[width=1in,height=1.25in,clip,keepaspectratio]{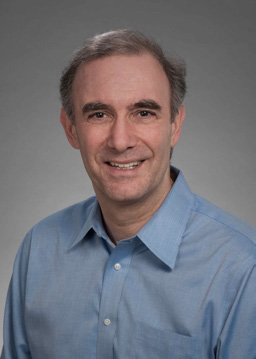}}]{Daniel Krischen}
(M'86-SM'91-F'07) received the degree in electrical and mechanical engineering from the Universite Libre de Bruxelles, Brussels, Belgium, and the M.Sc. and Ph.D. degrees from the University of Wisconsin, Madison, WI, USA, 1979, 1980, and 1985, respectively. \\
He is currently a Close Professor of Electrical Engineering with the University of Washington, Seattle, WA, USA. His research interests include smart grids, the integration of renewable energy sources in the grid, power system economics, and power system security.
\end{IEEEbiography}

\begin{IEEEbiography}[{\includegraphics[width=1in,height=1.25in,clip,keepaspectratio]{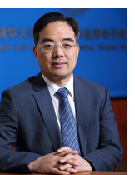}}]{Chongqing Kang}
(M'01-SM'08-F'17) received the Ph.D. degree from the Department of Electrical Engineering in Tsinghua University, Beijing, China, in 1997. \\
He is currently a Professor in Tsinghua University. His research interests include power system planning, power system operation, renewable energy, low carbon electricity technology and load forecasting.
\end{IEEEbiography}

\ifCLASSOPTIONcaptionsoff
  \newpage
\fi
\end{document}